\documentclass[%
reprint,
showpacs,preprintnumbers,
amsmath,amssymb,
aps,
prb,
]{revtex4-1}

\usepackage{graphicx}
\usepackage{dcolumn}
\usepackage{bm}
\usepackage{colortbl}

\begin{document}


\title{Kinetics of local $"$magnetic$"$ moment and non-stationary spin-polarized current in the single impurity Anderson-model}

\author{N.\,S.\,Maslova$^{1}$}
\altaffiliation{}
\author{V.\,N.\,Mantsevich$^{1}$}
\email{vmantsev@gmail.com} \altaffiliation{}
\author{P.\,I.\,Arseyev$^{2,3}$}

\affiliation{$^{1}$Moscow State University, Department of  Physics,
119991 Moscow, $^{2}$Russia P.N.Lebedev Physical Institute of RAS,
119991 Moscow, $^{3}$Russia National Research University Higher
School of Economics, Moscow, Russia
}%

\date{\today }
\begin{abstract}
We perform theoretical investigation of the localized state dynamics
in the presence of interaction with the reservoir and Coulomb
correlations. We analyze kinetic equations for electron occupation
numbers with different spins taking into account high order
correlation functions for the localized electrons. We reveal that in
the stationary state electron occupation numbers with the opposite
spins always have the same value - the stationary state is a
$"$paramagnetic$"$ one. $"$Magnetic$"$ properties can appear only in
the non-stationary characteristics of the single-impurity Anderson
model and in the dynamics of the localized electrons second order
correlation functions. We found, that for deep energy levels and
strong Coulomb correlations, relaxation time for initial
$"$magnetic$"$ state can be several orders larger than for
$"$paramagnetic$"$ one. So, long-living $"$magnetic$"$ moment can
exist in the system. We also found non-stationary spin polarized
currents flowing in opposite directions for the different spins in
the particular time interval.
\end{abstract}

 \keywords{} \maketitle

\section{Introduction}
The creation, diagnostics and controlled manipulation of charge and
spin states of the impurity atoms or quantum dots (QDs) is one of
the most important problems in nano-electronics now a days
\cite{Jacak98},\cite{Wiel02}, \cite{Arseyev},
\cite{Hornberger},\cite{Fransson}. Modern ultra small size
electronic devices design with a given set of electronic transport
parameters requires careful analysis of non-stationary effects,
transient processes and time evolution of electronic states prepared
at the initial time moment
\cite{Bar-Joseph},\cite{Gurvitz_1},\cite{Gurvitz_2},\cite{Arseyev_1},\cite{Stafford_1},\cite{Hazelzet},\cite{Cota}.
So, it is necessary to investigate the time dependent dynamics of
initial spin and charge configurations of correlated impurity or QD.
Moreover, the characteristics of stationary state of single impurity
interacting with the reservoir in the presence of strong Coulomb
correlations are not completely understood
\cite{Arseyev_2},\cite{Contreras-Pulido},\cite{Elste},\cite{Kennes}.

The possibility of the localized non-zero magnetic moment existence
on the single impurity or single-level QD, interacting with the
reservoir, in the absence of external magnetic field is still
unclear. Results obtained in the mean-field approximation for the
one-level Anderson model allowing the presence of magnetic state
(electron occupation numbers with opposite spins have different
values) for the single impurity with strong on-cite Coulomb
repulsion seems to be rather questionable.

The single-impurity Anderson model for a long time served as a basic
one for the understanding of the nature of local magnetic moments in
solids \cite{Anderson},\cite{Lieb}.  For a single partly occupied
impurity state, the correlation energy acts to prevent the
appearance of a non-vanishing ground-state spin, while in
low-density limit the Hartree-Fock theory still predicts a non-zero
magnetic moment over a range of parameters \cite{Schrierrer}. As it
was argued in \cite{Schrierrer} the magnetism is possible only when
several degenerate orbitals are present on the impurity in the
Anderson model. Local moment approach to the Anderson model has been
applied for the case of half-filling in \cite{Logan}.

The most adequate approach for this problem analysis is based on the
non-stationary kinetic equations for localized electron occupation
numbers and their correlation functions, taking into account all
high-order correlation functions for the localized electrons. The
simplest way to obtain the system of kinetic equations is the
Heisenberg approach. These equations can be also obtained by means
of Keldysh diagram technique, but it is more cumbersome procedure
\cite{Keldysh}.

In this paper we analyze the localized state dynamics in the
presence of interaction with the reservoir and Coulomb correlations
by means of kinetic equations for electron occupation numbers with
the different spins and second order correlation functions of the
localized electrons. We demonstrate that $"$magnetic$"$ state can be
distinguished from the $"$paramagnetic$"$ one be means of the
analysis of the non-stationary characteristics and dynamics of
second order correlation functions.

\section{Theoretical model and main results}

We consider non-stationary processes in the system of the
single-level impurity coupled to an electronic reservoir with
Coulomb interaction of the localized electrons. The model
Hamiltonian has the form:

\begin{eqnarray}
\hat{H}&=&\sum_{\sigma}\varepsilon_{1}\hat{n}_{1\sigma}+\sum_{k\sigma}\varepsilon_{k}\hat{c}_{k\sigma}^{+}\hat{c}_{k\sigma}+\nonumber\\
&+&U\hat{n}_{1\sigma}\hat{n}_{1-\sigma}+
\sum_{k\sigma}t_{k}(\hat{c}_{k\sigma}^{+}\hat{c}_{1\sigma}+\hat{c}_{1\sigma}^{+}\hat{c}_{k\sigma}).
\end{eqnarray}

Index $k$ labels continuous spectrum states in the lead, $t_{k}$-
tunneling transfer amplitude between the continuous spectrum states
and localized state with the energy $\varepsilon_1$ which is
considered to be independent of momentum and spin. Operators
$\hat{c}_{k}^{+}/\hat{c}_{k}$ correspond to the electrons
creation/annihilation in the continuous spectrum states $k$.
$\hat{n}_{1\sigma(-\sigma)}=\hat{c}_{1\sigma(-\sigma)}^{+}\hat{c}_{1\sigma(-\sigma)}$-localized
state electron occupation numbers, where operator
$\hat{c}_{1\sigma(-\sigma)}$ destroys electron with spin
$\sigma(-\sigma)$ on the energy level $\varepsilon_1$. $U$ is the
on-site Coulomb repulsion for the double occupation of the localized
state.

Our investigations deal with the low temperature regime when Fermi
level is well defined and the temperature is much lower than all the
typical energy scales in the system. Consequently the distribution
function of electrons in the leads (band electrons) is close to the
Fermi step.

Let us consider $\hbar=1$ elsewhere, so the motion equation for the
electron operators products
$\hat{c}_{1\sigma}^{+}\hat{c}_{1\sigma}$,
$\hat{c}_{1\sigma}^{+}\hat{c}_{k\sigma}$ and
$\hat{c}_{k^{'}\sigma}^{+}\hat{c}_{k\sigma}$ can be written as:

\begin{eqnarray}
i\frac{\partial \hat{c}_{1\sigma}^{+}\hat{c}_{1\sigma}}{\partial
t}&=&-\sum_{k,\sigma}t_{k}\cdot(\hat{c}_{k\sigma}^{+}\hat{c}_{1\sigma}-\hat{c}_{1\sigma}^{+}\hat{c}_{k\sigma}),\nonumber\\
i\frac{\partial \hat{c}_{1\sigma}^{+}\hat{c}_{k\sigma}}{\partial
t}&=&-(\varepsilon_1-\varepsilon_k)\cdot
\hat{c}_{1\sigma}^{+}\hat{c}_{k\sigma}-U\hat{n}_{1-\sigma}\cdot
\hat{c}_{1\sigma}^{+}\hat{c}_{k\sigma}+\nonumber\\
&+&t_{k}\cdot (\hat{n}_{1\sigma}-\hat{n}_{k\sigma}) -\sum_{k^{'}\neq
k}t_{k^{'}}\hat{c}_{k^{'}\sigma}^{+}\hat{c}_{k\sigma} \label{2}
\end{eqnarray}

and

\begin{eqnarray}
i\frac{\partial \hat{c}_{k^{'}\sigma}^{+}\hat{c}_{k\sigma}}{\partial
t}&=&-(\varepsilon_{k^{'}}-\varepsilon_k)\cdot
\hat{c}_{k^{'}\sigma}^{+}\hat{c}_{k\sigma}-\nonumber\\&-&t_{k^{'}}\cdot
\hat{c}_{1\sigma}^{+}\hat{c}_{k\sigma}+t_{k}\cdot
\hat{c}_{k^{'}\sigma}^{+}\hat{c}_{1\sigma}, \label{3}
\end{eqnarray}

where $\hat{n}_{k}=\hat{c}_{k\sigma}^{+}\hat{c}_{k\sigma}$ is an
occupation operator for the electrons in the reservoir. From
Eq.(\ref{3}) one can obtain:

\begin{eqnarray}
\sum_{k^{'}\neq
k}\hat{c}_{k^{'}\sigma}^{+}\hat{c}_{k\sigma}t_{k^{'}}&=&i\sum_{k^{'}}\int^{t}
dt_{1}\times\nonumber\\ \times [t_{k^{'}}^{2}
\hat{c}_{1\sigma}^{+}\hat{c}_{k\sigma}&-&t_{k}t_{k^{'}}
\hat{c}_{k^{'}\sigma}^{+}\hat{c}_{1\sigma}]\cdot e^{i\cdot(\varepsilon_{k}-\varepsilon_k^{'})\cdot(t-t_{1})}\label{4}.\nonumber\\
\end{eqnarray}

Combining Eqs. (\ref{4}) and (\ref{2}) we obtain:

\begin{eqnarray}
i\frac{\partial \hat{c}_{1\sigma}^{+}\hat{c}_{k\sigma}}{\partial
t}&+&[\varepsilon_1-\varepsilon_k+i\Gamma_{k}]\cdot\hat{c}_{1\sigma}^{+}\hat{c}_{k\sigma}+U\hat{n}_{1-\sigma}\cdot
\hat{c}_{1\sigma}^{+}\hat{c}_{k\sigma}=\nonumber\\&=&t_{k}\cdot(\hat{n}_{1\sigma}-\hat{n}_{k\sigma})+i\sum_{k^{'}}\int^{t}
dt_{1}\times\nonumber\\ &\times& t_{k}t_{k^{'}}\hat{c}_{k^{'}\sigma}^{+}\hat{c}_{1\sigma}\cdot e^{i\cdot(\varepsilon_{k}-\varepsilon_k^{'})\cdot(t-t_{1})},\nonumber\\
\label{5}
\end{eqnarray}

where $\Gamma_{k}=\nu_{k0}t_{k(p)}^{2}$, $\nu_{k0}$ - is the
unperturbed density of states in the tunneling contact lead.
Multiplying Eq. (\ref{5}) by electron operators
$(1-\hat{n}_{1-\sigma})$ and $\hat{n}_{1-\sigma}$ we obtain the
following expressions:

\begin{eqnarray}
(1-\hat{n}_{1-\sigma})&\cdot& i\frac{\partial
\hat{c}_{1\sigma}^{+}\hat{c}_{k\sigma}}{\partial
t}+[\varepsilon_1-\varepsilon_k+i\Gamma_{k}][1-\hat{n}_{1-\sigma}]\hat{c}_{1\sigma}^{+}\hat{c}_{k\sigma}=\nonumber\\&=&(1-\hat{n}_{1-\sigma})\cdot
[t_{k}\cdot(\hat{n}_{1\sigma}-\hat{n}_{k\sigma})+\nonumber\\&+&i\sum_{k^{'}}\int^{t}
dt_{1}t_{k}t_{k^{'}}\hat{c}_{k^{'}\sigma}^{+}(t_{1})\hat{c}_{1\sigma}(t_{1})e^{i\cdot(\varepsilon_{k}-\varepsilon_k^{'})\cdot(t-t_{1})}],\nonumber\\
\hat{n}_{1-\sigma}&\cdot& i\frac{\partial
\hat{c}_{1\sigma}^{+}\hat{c}_{k\sigma}}{\partial
t}+[\varepsilon_1-\varepsilon_k+U+i\Gamma_{k}]\hat{n}_{1-\sigma}\hat{c}_{1\sigma}^{+}\hat{c}_{k\sigma}=\nonumber\\&=&\hat{n}_{1-\sigma}\cdot
[t_{k}\cdot(\hat{n}_{1\sigma}-\hat{n}_{k\sigma})+\nonumber\\&+&i\sum_{k^{'}}\int^{t}
dt_{1}t_{k}t_{k^{'}}\hat{c}_{k^{'}\sigma}^{+}(t_{1})\hat{c}_{1\sigma}(t_{1})e^{i\cdot(\varepsilon_{k}-\varepsilon_k^{'})\cdot(t-t_{1})}].\nonumber\\
\label{6}
\end{eqnarray}

If condition $\frac{\varepsilon_1-\varepsilon_F}{\Gamma}>>1$ is
fulfilled, $\hat{n}_{1-\sigma}$ is a slowly varying variable in
comparison with the $\hat{c}_{1\sigma}^{+}\hat{c}_{k\sigma}$
($\frac{\partial}{\partial
t}\hat{n}_{1-\sigma}<<\frac{\partial}{\partial
t}\hat{c}_{1\sigma}^{+}\hat{c}_{k\sigma}$). Consequently, it is
reasonable to consider that:

\begin{eqnarray}
\frac{\partial}{\partial
t}\hat{n}_{1-\sigma}\hat{c}_{1\sigma}^{+}\hat{c}_{k\sigma}\sim
\hat{n}_{1-\sigma}\frac{\partial}{\partial
t}\hat{c}_{1\sigma}^{+}\hat{c}_{k\sigma}.\label{7}
\end{eqnarray}

So, terms $(\frac{\partial}{\partial
t}\hat{n}_{1-\sigma})\hat{c}_{1\sigma}^{+}\hat{c}_{k\sigma}$  are
omitted. Omitted terms $\frac{\partial \hat{n}_{1-\sigma}}{\partial
t}\hat{c}_{1\sigma}^{+}\hat{c}_{k\sigma}$ in the right hand side of
Eq. (\ref{6}) are responsible for the Kondo effect.

One can get expressions for
$(1-\widehat{n}_{1-\sigma})\hat{c}_{1\sigma}^{+}\hat{c}_{k\sigma}$
and $\widehat{n}_{1-\sigma}\hat{c}_{1\sigma}^{+}\hat{c}_{k\sigma}$
(applying the procedure similar to the one which was used to obtain
Eq.(\ref{4}) from Eq.(\ref{3})) and then for
$\hat{c}_{1\sigma}^{+}\hat{c}_{k\sigma}$.

Substituting expression for $\hat{c}_{1\sigma}^{+}\hat{c}_{k\sigma}$
to Eq. (\ref{2}) we obtain equations, which determine time evolution
of electron occupation numbers $\hat{n}_{1\sigma}$. It is necessary
to note, that the last term in Eq. (\ref{5}) after summation over
index $k$ doesn't contribute to the non-stationary equations for the
electron occupation numbers $\hat{n}_{1\sigma}$. So, the time
evolution of the electron occupation numbers operators are governed
by the following system of equations:

\begin{eqnarray}
\frac{\hat{n}_{1\sigma}}{\partial
t}=-2\Gamma_{k}[\hat{n}_{1\sigma}-(1-\hat{n}_{1-\sigma})
\hat{N}_{k\varepsilon}^{\sigma}(t)-\hat{n}_{1-\sigma}
\hat{N}_{k\varepsilon+U}^{\sigma}(t)],\nonumber\\
\frac{\hat{n}_{1-\sigma}}{\partial
t}=-2\Gamma_{k}[\hat{n}_{1-\sigma}-(1-\hat{n}_{1\sigma})
\hat{N}_{k\varepsilon}^{-\sigma}(t)-\hat{n}_{1\sigma}
\hat{N}_{k\varepsilon+U}^{-\sigma}(t)]. \label{8}\nonumber\\
\end{eqnarray}

Before we define $\hat{N}_{k\varepsilon}^{\sigma}(t)$ and
$\hat{N}_{k\varepsilon+U}^{\sigma}(t)$, one necessary explanation
should be made. One can see, that after applying the approximation
(\ref{7}) to Eq.(\ref{6}) the omitted terms are of the order of
$\Gamma_k$. It means, that parameter $\Gamma_k$ in Eq.(\ref{6})
should be replaced by effective parameter $\Gamma\sim\Gamma_k$.

Operators $\hat{N}_{k\varepsilon}^{\sigma(-\sigma)}(t)$ and
$\hat{N}_{k\varepsilon+U}^{\sigma(-\sigma)}(t)$ in Eq.(\ref{8}) are
defined as:

\begin{eqnarray}
\hat{N}_{k\varepsilon}^{\sigma}(t)=\hat{N}_{k\varepsilon}^{-\sigma}(t)=\frac{1}{2}i
\int
d\varepsilon_{k}\hat{n}_{k}^{\sigma}(\varepsilon_{k})\times\nonumber\\ \times[\frac{1-e^{i(\varepsilon_1+i\Gamma-\varepsilon_{k})t}}{\varepsilon_1+i\Gamma-\varepsilon_{k}}-\frac{1-e^{-i(\varepsilon_1-i\Gamma-\varepsilon_{k})t}}{\varepsilon_1-i\Gamma-\varepsilon_{k}}],\nonumber\\
\hat{N}_{k\varepsilon+U}^{\sigma}(t)=\hat{N}_{k\varepsilon+U}^{-\sigma}(t)=\frac{1}{2}i
\int
d\varepsilon_{k}\hat{n}_{k}^{\sigma}(\varepsilon_{k})\times\nonumber\\
\times[\frac{1-e^{i(\varepsilon_1+U+i\Gamma-\varepsilon_{k})t}}{\varepsilon_1+U+i\Gamma-\varepsilon_{k}}-\frac{1-e^{-i(\varepsilon_1+U-i\Gamma-\varepsilon_{k})t}}{\varepsilon_1+U-i\Gamma-\varepsilon_{k}}],\nonumber\\
\label{N_01}\end{eqnarray}

Further we'll consider the situation when the reservoir is
paramagnetic:
$\hat{N}_{k\varepsilon}^{\sigma}(t)=\hat{N}_{k\varepsilon}^{-\sigma}(t)=\hat{N}_{k\varepsilon}(t)$
and
$\hat{N}_{k\varepsilon+U}^{\sigma}(t)=\hat{N}_{k\varepsilon+U}^{-\sigma}(t)=\hat{N}_{k\varepsilon+U}(t)$.
We can obtain equations for the occupation numbers of localized
electrons $n_{1\pm\sigma}$  by averaging Eqs. (\ref{8})-(\ref{N_01})
for the operators and by decoupling electrons occupation numbers in
the reservoir. Such decoupling procedure is reasonable if one
considers that electrons in the macroscopic reservoir is in the
thermal equilibrium. After decoupling one has to replace electron
occupation numbers operators in the reservoir $\hat{n}_{k}^{\sigma}$
by the Fermi distribution functions $f_{k}^{\sigma}$ in Eqs.
(\ref{5})-(\ref{N_01}).

We'll investigate time dependent dynamics of the electron occupation
numbers and their correlation functions for the different initial
conditions: 1) the non-zero localized magnetic moment exists on the
impurity ($|n_{1\sigma}-n_{1-\sigma}|\sim1$). Such state can be
prepared due to the applied external magnetic field $\mu
B>>\varepsilon_1$, which is switched $"$off$"$ at the initial time
moment $t=0$; 2) the initial state close to highly occupied
paramagnetic one ($|1-n_{1\pm\sigma}|<<1$) can be prepared by the
applied bias voltage $|eV|>\varepsilon_1+U$ switching $"$off$"$ or
$"$on$"$ at the initial time moment $t=0$; 3) the initial state
close to the low occupied paramagnetic one ($|n_{1\pm\sigma}|<<1$)
can be prepared by the applied bias voltage $|eV|<\varepsilon_1$
switching $"$off$"$ or $"$on$"$ at the initial time moment $t=0$. It
will be shown that relaxation time scale strongly depends on the
properties of the initially prepared state. The solution of kinetic
equations (\ref{8}) for electron occupation numbers $n_{\pm\sigma}$
can be easily found numerically for the arbitrary initial
conditions.

If one is interested in the system evolution for the time scales
$t>>\frac{1}{\varepsilon_1}$, fast oscillating terms, which contain
time dependent exponents can be neglected. Consequently, functions
$N_{k\varepsilon}$ and $N_{k\varepsilon+U}$ become time-independent.
So, localized electrons occupation numbers $n_{1\sigma}$,
$n_{1-\sigma}$ satisfy the linear system of equations, which can be
easily solved for the arbitrary initial conditions:

\begin{eqnarray}
n_{1\sigma}&=&\frac{N_{k\varepsilon}}{1+\Delta
N}\cdot(1-e^{\lambda_2t})+\nonumber\\&+&\frac{n_{1\sigma}(0)-n_{1-\sigma}(0)}{2}\cdot
e^{\lambda_1t}+\frac{n_{1\sigma}(0)+n_{1-\sigma}(0)}{2}\cdot
e^{\lambda_2t},\nonumber\\
n_{1-\sigma}&=&\frac{N_{k\varepsilon}}{1+\Delta
N}\cdot(1-e^{\lambda_2t})+\nonumber\\&+&\frac{n_{1-\sigma}(0)-n_{1\sigma}(0)}{2}\cdot
e^{\lambda_1t}+\frac{n_{1-\sigma}(0)+n_{1\sigma}(0)}{2}\cdot
e^{\lambda_2t}.\nonumber\\
\end{eqnarray}

The eigenvalues $\lambda_{1,2}$ are determined as:

\begin{eqnarray}
\lambda_{1,2}=-2\Gamma\cdot(1\mp\Delta N)
\end{eqnarray}

and

\begin{eqnarray}
\Delta N=N_{k\varepsilon}-N_{k\varepsilon+U}.
\end{eqnarray}

Straightforward calculations yield:

\begin{eqnarray}
\Delta
N=\frac{1}{\pi}[\arctan(-\frac{\varepsilon_1}{\Gamma})-\arctan(\frac{-\varepsilon_1+W}{\Gamma})-\nonumber\\-
\arctan(-\frac{\varepsilon_1+U}{\Gamma})+\arctan(\frac{W-(\varepsilon_1+U)}{\Gamma})].
\end{eqnarray}

where $W$ - is a band width for the conduction electrons in the
reservoir.

In the case of the large bandwidth for $\varepsilon_1<0$,
$\varepsilon_1+U>0$, $|\varepsilon_1|/\Gamma>>1$ and
$(\varepsilon_1+U)/\Gamma>>1$ one can obtain:

\begin{eqnarray}
\frac{|\lambda_{1}|}{2\Gamma}&\sim&\frac{\Gamma
U}{2|\varepsilon_1|(U-|\varepsilon_1|)},\nonumber\\
\frac{|\lambda_{2}|}{2\Gamma}&\sim&2-\frac{\Gamma
U}{2|\varepsilon_1|(U-|\varepsilon_1|)}.
\end{eqnarray}

\begin{figure}
\includegraphics[width=60mm]{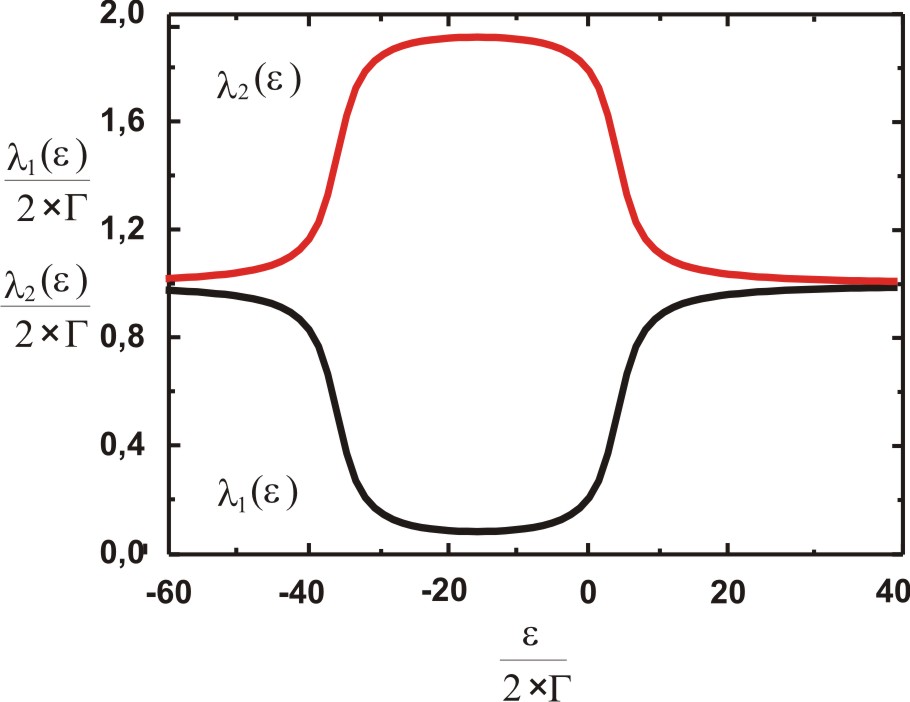}%
\caption{(Color online) Normalized relaxation rates
$|\lambda_{1,2}(\varepsilon)|/2\Gamma$ as a functions of the
localized state energy level position $\varepsilon/2\Gamma$ for
$U/2\Gamma=7.5$ and $\Gamma=1$. } \label{figure1}
\end{figure}

\begin{figure}
\includegraphics[width=60mm]{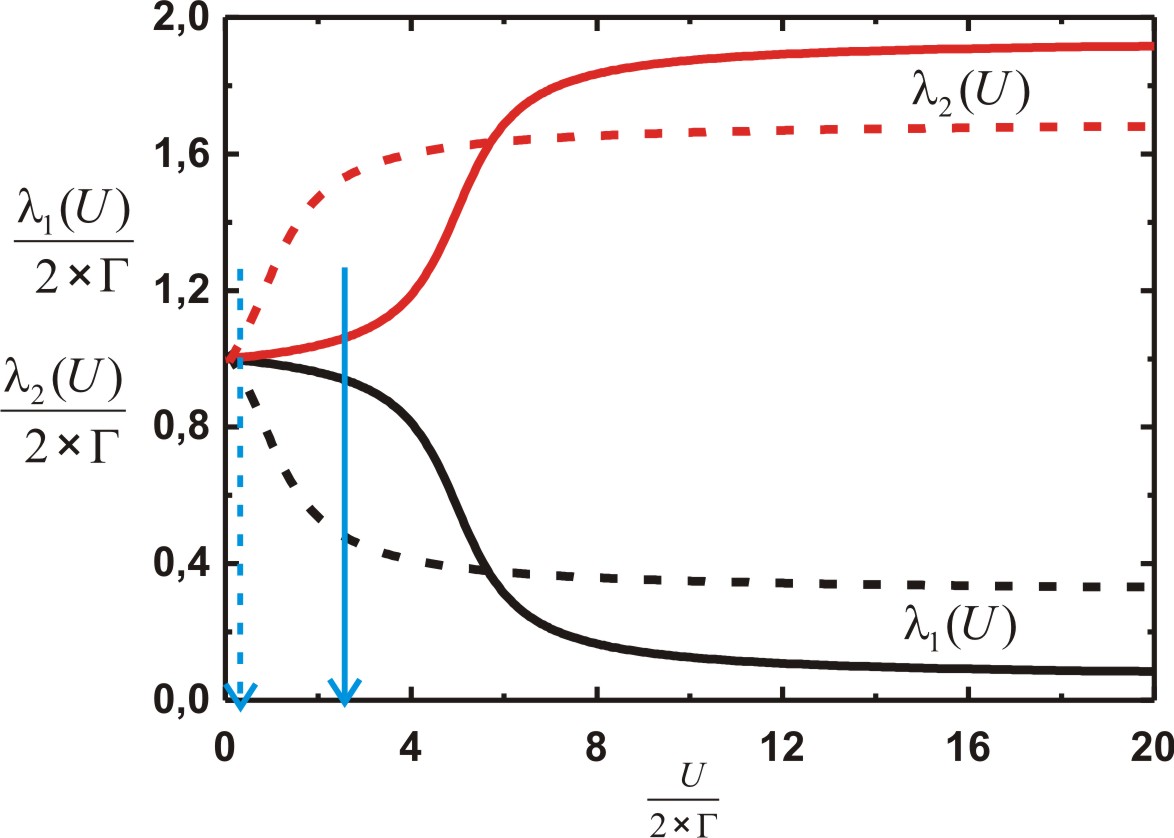}%
\caption{(Color online) Normalized relaxation rates
$|\lambda_{1,2}(U)|/2\Gamma$ as a functions of Coulomb interaction
value $U/2\Gamma$ for $\Gamma=1$. Blue arrows demonstrate localized
states energy levels values. Solid lines
$|\varepsilon|/2\Gamma=2.5$; dashed lines
$|\varepsilon|/2\Gamma=0.375$.} \label{figure2}
\end{figure}

For the large values of Coulomb interaction and deep energy level of
the localized state, relaxation time $|\lambda_{1}|^{-1}$ can be
several orders larger than the relaxation time of initial localized
state in the absence of Coulomb interaction. Relaxation rates
behavior for the different system parameters is shown in
Fig.\ref{figure1} and Fig.\ref{figure2}. Relaxation rates for the
magnetic moment and charge strongly differ for the deep energy
levels (see Fig.\ref{figure1}), but they  nearly become equal for
the energy levels $\varepsilon_1>0$ or $\varepsilon_1+U<0$. The role
of Coulomb correlations was also analyzed (see Fig.\ref{figure2}).
The presence of Coulomb correlations results in the increasing of
the relaxation values difference.

Typical time for the system to achieve the stationary state depends
on the initial conditions. For the $"$paramagnetic$"$ initial
conditions ($n_{1\sigma}(0)=n_{1-\sigma}(0)$) relaxation rate to the
stationary state is determined by $|\lambda_2|=2\Gamma\cdot(1+\Delta
N)$ and in the case of the $"$magnetic$"$ initial conditions
($|n_{1\sigma}(0)-n_{1-\sigma}(0)|\sim1$) relaxation rate to the
stationary state is determined by $|\lambda_1|=2\Gamma\cdot(1-\Delta
N)$. Consequently, we have long living $"$magnetic$"$ moments.

In the presence of interaction with the $"$paramagnetic$"$ reservoir
($\Gamma\neq0$) stationary state is always a $"$paramagnetic$"$ one:

\begin{eqnarray}
n_{1}^{st}=n_{1\sigma}=n_{1-\sigma}=\frac{N_{k\varepsilon}}{1+\Delta
N}.
\end{eqnarray}

The behavior of localized state electron occupation numbers for the
different initial conditions and the set of system parameters is
depicted in Fig.(\ref{figure4})-Fig.(\ref{figure5}). Panels a,c
correspond to the case when Coulomb interaction is present and
panels b,d - describe the situation when relaxation takes place in
the absence of Coulomb correlations. Magnetic properties are
revealed for $|\varepsilon_1|<0$, $\varepsilon_1+U>0$,
$|\varepsilon_1|/\Gamma>>1$ and $(\varepsilon_1+U)/\Gamma>>1$ in the
slow relaxation of the initial $"$magnetic$"$ state, prepared at
$t=t_0$ ($|\lambda_1|<<|\lambda_2|$) (see Fig.\ref{figure4}).
Nonzero magnetic moment is present on the impurity for
$t>>(2\Gamma)^{-1}$. So, the time scale when magnetic moment exists
on the impurity (see panel a,c in
Fig.(\ref{figure4})-Fig.(\ref{figure5})) strongly exceeds the
relaxation time for the impurity state without Coulomb interaction
(see panel b,d in Fig.(\ref{figure4})-Fig.(\ref{figure5})). Obtained
results demonstrate, that the stationary state of the single
impurity with Coulomb correlations in the presence of interaction
with the reservoir is always $"$paramagnetic$"$. The mean values of
the electron occupation numbers with the opposite spin directions in
the stationary case have the same magnitudes for any value of the
on-site Coulomb repulsion, contrary to the results obtained in the
mean-field approximation.

We revealed that typical times of the stationary state formation are
determined by the initial conditions. For the deep energy levels and
strong Coulomb correlations (see panels a in the
Fig.(\ref{figure4})-Fig.(\ref{figure5}), relaxation time for the
initial $"$magnetic$"$ state can be several orders larger than for
the $"$paramagnetic$"$ one. This fact reflects the $"$magnetic$"$
nature of the single occupied localized state with strong Coulomb
correlations. The presence of long-living $"$magnetic$"$ moment
depends on the ratio between the system parameters: the single
electron level position, the value of Coulomb interaction and
coupling to reservoir.

Non-stationary spin polarized currents flowing in opposite
directions for different spins exists in the system in the
particular time interval (see Fig.\ref{figure4}). Non-stationary
spin-polarized tunneling currents are determined by the right-hand
side of Eq. (\ref{8}). For $|\lambda_1|^{-1}>t>(2\cdot\Gamma)^{-1}$:

\begin{eqnarray}
\frac{1}{e}\cdot I^{\pm}=\frac{\partial n_{1\pm\sigma}}{\partial
t}\sim\pm [n_{1\sigma}(0)-n_{1-\sigma}(0)]\cdot\lambda_1\cdot
e^{\lambda_1t}.\nonumber\\
\end{eqnarray}

\begin{figure}
\includegraphics[width=60mm]{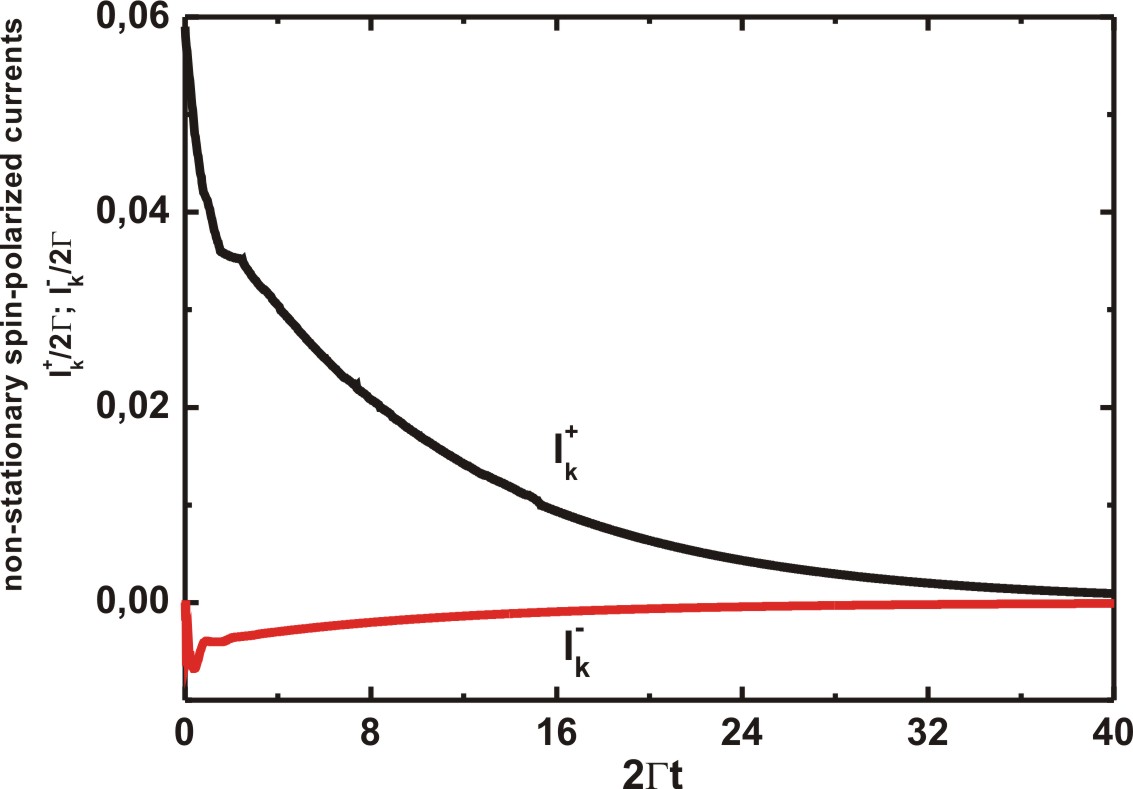}%
\caption{(Color online) Normalized non-stationary spin-polarized
tunneling currents $I^{+}(t)/2\Gamma$ (black line) and
$I^{-}(t)/2\Gamma$ (red line). $n_{1\sigma}(0)=1$ and
$n_{1-\sigma}(0)=0$; $\varepsilon/2\Gamma=-2.5$; $U/2\Gamma=7.5$ and
$\Gamma=1$.} \label{figure3}
\end{figure}

\begin{figure}
\includegraphics[width=80mm]{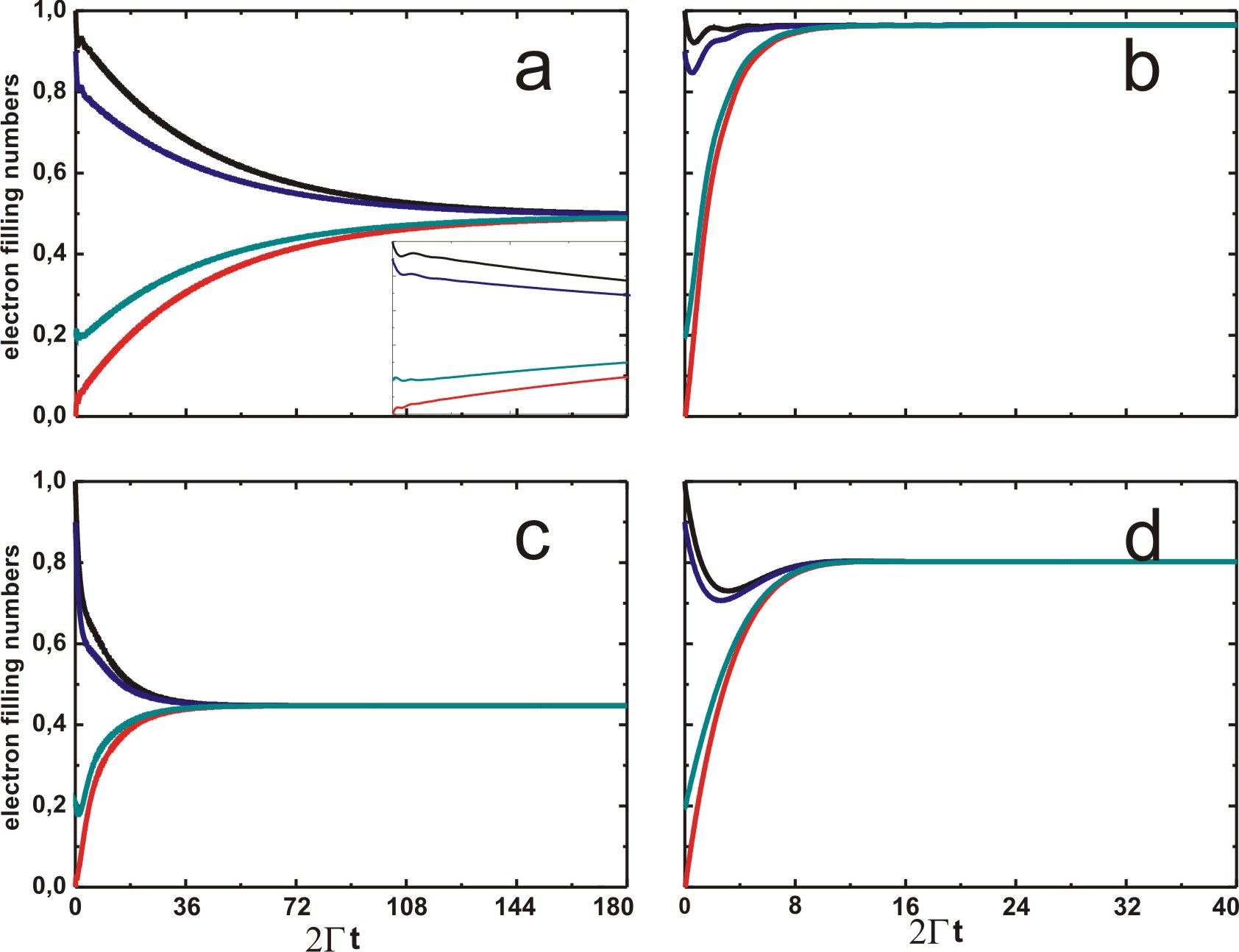}%
\caption{(Color online) Electron occupation numbers time evolution
for the $"$magnetic$"$ initial conditions. Black and blue lines
demonstrate $n_{1\sigma}(t)$, red and green lines -
$n_{1-\sigma}(t)$. a),b) $\varepsilon/2\Gamma=-2.5$; c),d)
$\varepsilon/2\Gamma=-0.375$. a).,c). long-living $"$magnetic$"$
moments in the presence of Coulomb interaction $U/2\Gamma=7.5$;
b).,d). fast relaxation in the absence of Coulomb interaction
$U/2\Gamma=0$. Parameter $\Gamma=1$ is the same for all the figures.
Black line - $n_{1\sigma}(0)=1$, red line - $n_{1-\sigma}(0)=0$,
blue line - $n_{1\sigma}(0)=0.9$, green line -
$n_{1-\sigma}(0)=0.2$. Insert demonstrates the presence of
oscillations at the very beginning of charge relaxation.}
\label{figure4}
\end{figure}

\begin{figure}
\includegraphics[width=80mm]{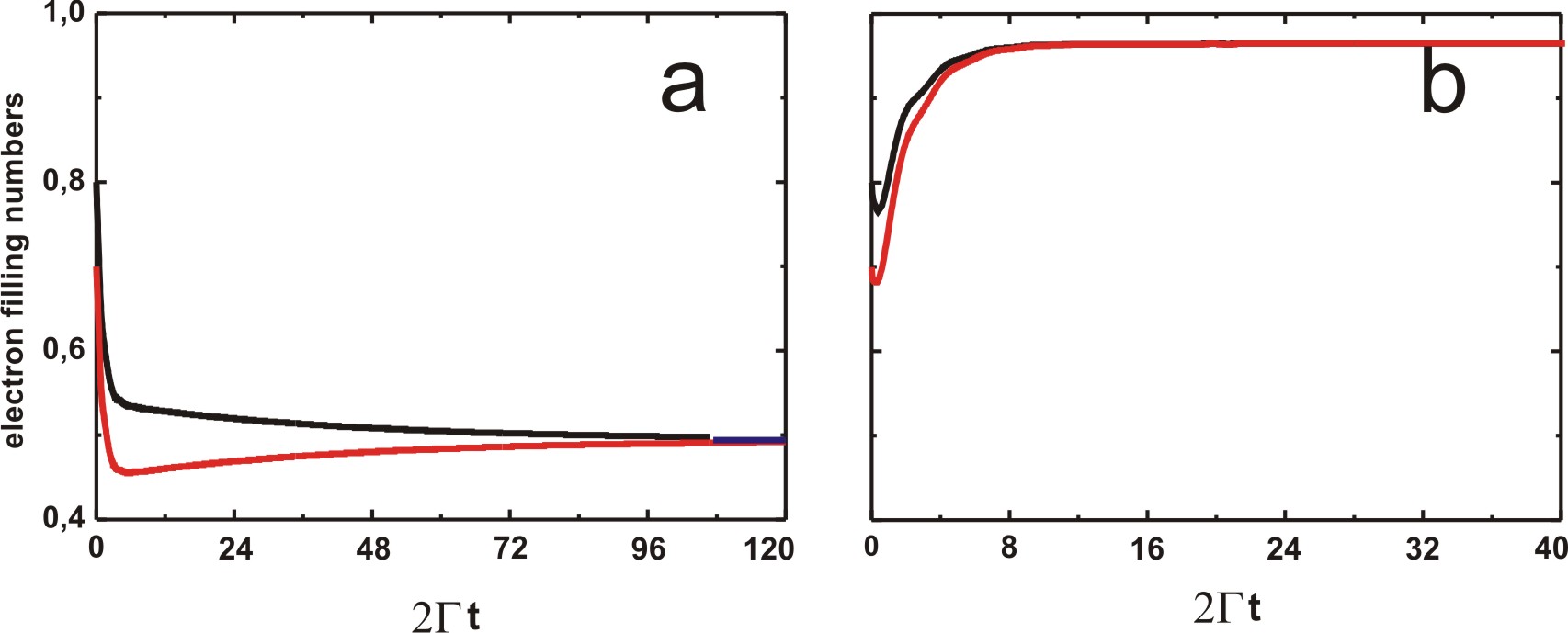}%
\caption{(Color online) The absence of large time scale in electron
occupation numbers time evolution for the initial conditions close
to the $"$paramagnetic$"$ one. Black line demonstrates
$n_{1\sigma}(t)$, red line - $n_{1-\sigma}(t)$. a). in the presence
of Coulomb interaction $U/2\Gamma=7.5$; b). in the absence of
Coulomb interaction $U/2\Gamma=0$. Parameters
$\varepsilon/2\Gamma=-2,5$ and $\Gamma=1$ are the same for all the
figures. Black line $n_{1\sigma}(0)=0.8$, red line
$n_{1-\sigma}(0)=0.7$.} \label{figure5}
\end{figure}

For typical $\Gamma\sim1\div10$ meV and $|\varepsilon|\sim50$ meV,
corresponding to the situation depicted in Fig.\ref{figure3} the
non-stationary spin-polarized current value is about $1\div10$ nA
($1nA\simeq6\times10^{9}e/sec$).

Charge transfer by the electrons with the $"$up$"$ and $"$down$"$
spins is determined as:

\begin{eqnarray}
\frac{1}{e}\cdot \Delta Q^{\pm}=
n_{1\pm\sigma}(0)-\frac{N_{k\varepsilon}}{1+\Delta N}.
\end{eqnarray}

For $n_{1\sigma}(0)-n_{1-\sigma}(0)\sim1$ and deep energy levels in
the presence of strong Coulomb interaction:

\begin{eqnarray}
|\Delta Q^{+}|-|\Delta Q^{-}|\sim\frac{\Gamma}{4\varepsilon_1}.
\end{eqnarray}

So, the total non-stationary charge transfer is connected with the
particular spin electrons, but it's value is small for
$\Gamma/\varepsilon_1<1$. This situation resembles the spin-Hall
systems with two types of $"$edge$"$ states with the opposite
velocities and spins at each system boundary with negligible charge
transfer from the one boundary to the another \cite{Hasan}.

\begin{figure}
\includegraphics[width=60mm]{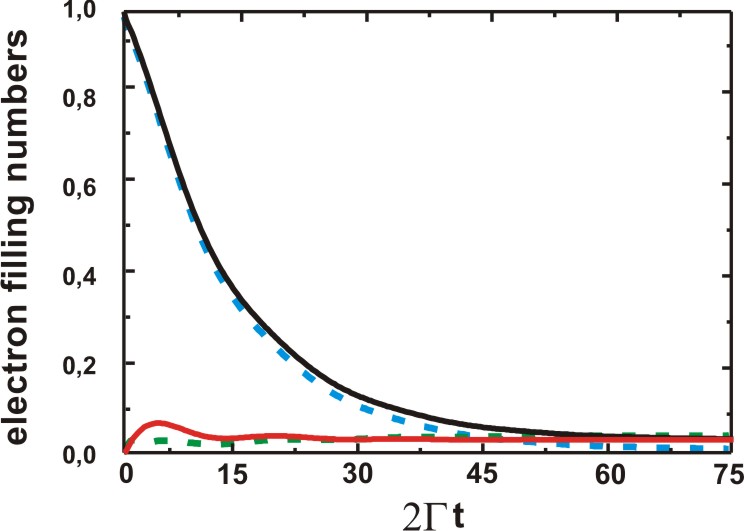}%
\caption{(Color online) Electron occupation numbers time evolution
for $\varepsilon/2\Gamma=2.5$, $\Gamma=1$. Black solid and blue
dashed lines demonstrate $n_{1\sigma}(t)$, red solid line and green
dashed lines - $n_{1-\sigma}(t)$. Solid lines $U/2\Gamma=7.5$,
dashed lines $U/2\Gamma=0$.} \label{figure6}
\end{figure}

If impurity energy level is localized above the Fermi level $E_F$,
two time scales $|\lambda_{1}|^{-1}$ and $|\lambda_{2}|^{-1}$ are of
the same order even for strong Coulomb interaction and for magnetic
initial conditions (see Fig.\ref{figure6}).

The magnetic properties can be also analyzed from the time
dependence of the stationary correlation functions for the electron
occupation numbers:

\begin{eqnarray}
K^{\sigma\sigma^{'}}(t-t^{'})=<n_{1\sigma}(t)n_{1\sigma^{'}}(t^{'})>.
\end{eqnarray}

Correlation functions $K^{\sigma\sigma^{'}}(\tau=t-t^{'})$ satisfy
the system of equations, which is derived from Eq.(\ref{8}) for
electron occupation numbers:

\begin{eqnarray}
\frac{\partial}{\partial t}K^{+-}&=&-2\Gamma_k[K^{+-}+\Delta
NK^{--}-N_{k\varepsilon}n_{1-\sigma}],\nonumber\\
\frac{\partial}{\partial t}K^{--}&=&-2\Gamma_k[K^{--}+\Delta
NK^{+-}-N_{k\varepsilon}n_{1-\sigma}]. \label{9}\end{eqnarray}

Initial conditions are determined as:

\begin{eqnarray}
K^{+-}(t,t)=K^{+-}(0)=\frac{N_{k\varepsilon+U}\cdot N_{k\varepsilon}}{1+\Delta N}\nonumber\\
K^{--}(0)=n_{1}^{st}=\frac{N_{k\varepsilon}}{1+\Delta N}.
\end{eqnarray}

Time evolution of the correlation functions can be obtained from the
Eq. (\ref{9}):

\begin{eqnarray}
K^{+-}(\tau)=\frac{N_{k\varepsilon}^{2}}{(1+\Delta
N)^{2}}\cdot[1-e^{\lambda_2\tau}]+\nonumber\\+\frac{N_{k\varepsilon}[N_{k\varepsilon+U}-1]}{2[1+\Delta
N]}\cdot
e^{\lambda_1\tau}+\frac{N_{k\varepsilon}[N_{k\varepsilon+U}+1]}{2[1+\Delta
N]}\cdot e^{\lambda_2\tau},\nonumber\\
K^{--}(\tau)=\frac{N_{k\varepsilon}^{2}}{(1+\Delta
N)^{2}}\cdot[1-e^{\lambda_2\tau}]+\nonumber\\+\frac{N_{k\varepsilon}[N_{k\varepsilon+U}+1]}{2[1+\Delta
N]}\cdot
e^{\lambda_2\tau}+\frac{N_{k\varepsilon}[1-N_{k\varepsilon+U}]}{2[1+\Delta
N]}\cdot e^{\lambda_1\tau}.\nonumber\\
\end{eqnarray}

\begin{figure}
\includegraphics[width=60mm]{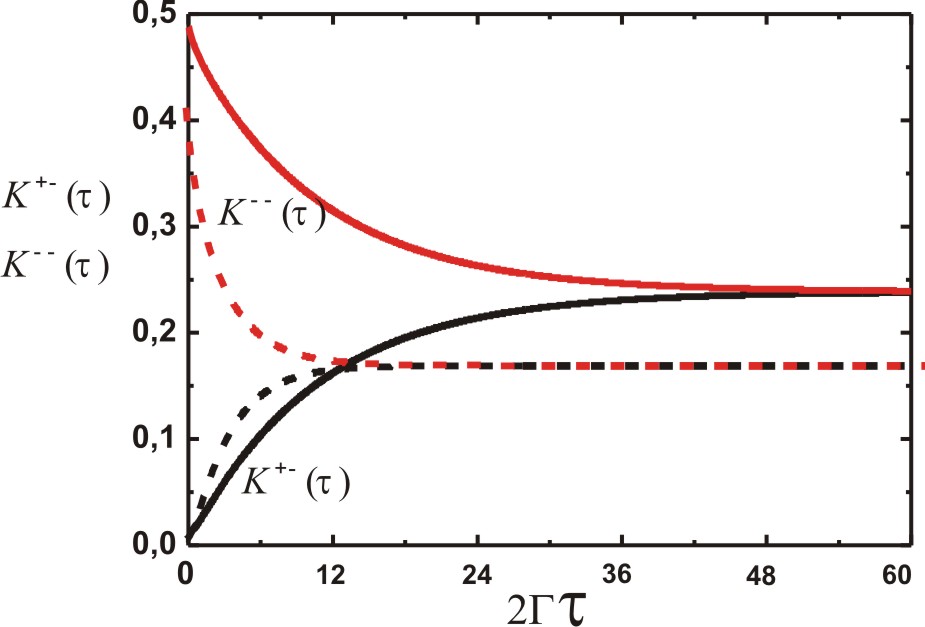}%
\caption{(Color online) Correlation functions time evolution for the
$"$magnetic$"$ initial conditions. Black lines demonstrate
$K^{+-}_{\tau}$, red lines - $K^{--}_{\tau}$. Solid lines
$\varepsilon/2\Gamma=-2.5$; Dashed lines
$\varepsilon/2\Gamma=-0.375$. Parameters $U/2\Gamma=7.5$ and
$\Gamma=1$.} \label{figure7}
\end{figure}

The behavior of the stationary correlation functions for the
localized electrons occupation numbers with the different spin
orientation is depicted in Fig.\ref{figure7}. It is clearly evident,
that for the deep energy levels correlation functions time evolution
is much lower, than for the states with shallow energy levels.
Autocorrelation function for the electron occupation numbers with
the opposite spins tends to zero for the strong Coulomb interaction.
Such behavior points to the possibility of the presence of non-zero
magnetic moment in a certain time interval.

For $\tau\rightarrow\infty$ correlation functions turns to the
product of the decoupled electronic occupation numbers mean values:

\begin{eqnarray}
K^{+-st}=K^{--st}\simeq(\frac{N_{k\varepsilon}}{1+\Delta}) .
\end{eqnarray}

So for $\tau<\frac{1}{|\lambda_1|}$ the $"$magnetic$"$ correlations
are still present in the system. Time evolution of $K^{+-}(\tau)$
and $K^{--}(\tau)$ is depicted in Fig.\ref{figure7}.

\section{Conclusion}
We demonstrated that the difference between $"$magnetic$"$ and
$"$paramagnetic$"$ states in the single-impurity Anderson model
appears only in the non-stationary characteristics of the system and
in the second order correlation functions behavior. Localized state
dynamics in the presence of interaction with the reservoir and
Coulomb correlations has been analyzed  by means of the kinetic
equations for the electron occupation numbers with the different
spins, taking into account high order correlation functions for the
localized electrons.

We revealed that the stationary state of the single impurity with
Coulomb correlations in the presence of interaction with the
reservoir is always a $"$paramagnetic$"$ one, even when interaction
is weak. Electron occupation numbers with the opposite spin in the
stationary case have are equal for any value of the on-site Coulomb
repulsion, contrary to the results obtained in the mean-field
approximation. To reveal $"$magnetic$"$ properties for the
single-impurity Anderson model one has to analyze non-stationary
system characteristics.

We showed that typical times of the stationary state formation
depend on the initial conditions. For the deep energy levels and
strong Coulomb correlations, relaxation time for the initial
$"$magnetic$"$ state can be several orders larger than for the
$"$paramagnetic$"$ one. This fact reflects the $"$magnetic$"$ nature
of the single occupied localized state with the strong Coulomb
correlations. Described relaxation times difference allows to
distinguish the $"$magnetic$"$ state on the localized impurity from
the $"$paramagnetic$"$ one. The existence of long-living
$"$magnetic$"$ moment depends on the ratio between the system
parameters: the single electron level position, the value of Coulomb
interaction and coupling to reservoir.

We analyzed the behavior of the correlation functions for the
localized electrons occupation numbers with the different spin
orientation. For the large time scales, which can strongly exceed
relaxation time of the system in the absence of Coulomb interaction,
rather strong correlations of the electron occupation numbers are
present. Such behavior of correlation functions points to the
existence of magnetic regime.

For initially magnetic impurities non-stationary spin polarized
currents flowing in the opposite directions for the different spins
exist in the system in the particular time interval similar to the
spin-Hall systems with the two types of the $"$edge$"$ states with
opposite velocities and spins at each boundary.

This work was supported by RFBR grant $16-32-60024$ $mol-a-dk$ and
by RFBR grant $14-02-00434$.

 \pagebreak

\end{document}